# Stego Quality Enhancement by Message Size Reduction and Fibonacci Bit-plane Mapping

Alan A. Abdulla, Harin Sellahewa and Sabah A. Jassim

Applied Computing Department, University of Buckingham, Buckingham, UK
alananwer@yahoo.com {harin.sellahewa; sabah.jassim}@buckingham.ac.uk

**Abstract.** An efficient 2-step steganography technique is proposed to enhance stego image quality and secret message un-detectability. The first step is a pre-processing algorithm that reduces the size of secret images without losing information. This results in improved stego image quality compared to other existing image steganography methods. The proposed secret image size reduction (SISR) algorithm is an efficient spatial domain technique. The second step is an embedding mechanism that relies on Fibonacci representation of pixel intensities to minimize the effect of embedding on the stego image quality. The improvement is attained by using bit-plane(s) mapping instead of bit-plane(s) replacement for embedding. The proposed embedding mechanism outperforms the binary based LSB randomly embedding in two ways: reduced effect on stego quality and increased robustness against statistical steganalysers. Experimental results demonstrate the benefits of the proposed scheme in terms of: 1) SISR ratio (indirectly results in increased capacity); 2) quality of the stego; and 3) robustness against steganalysers such as RS, and WS. Furthermore, experimental results show that the proposed SISR algorithm can be extended to be applicable on DICOM standard medical images. Future security standardization research is proposed that would focus on evaluating the security, performance, and effectiveness of steganography algorithms.

## 1 INTRODUCTION

Steganography is concerned with concealing a secret message by embedding it in another innocuous message during seemingly mundane communication sessions in such a way that only the sender and intended recipient are aware of the secret's existence. Steganography is becoming a common tool in protecting sensitive communications used by: intelligence and law enforcing agencies to prevent crime and terrorism; in health care systems to maintain the privacy of critical information such as medical records; and in financial organizations such as banks to prevent customers' account information from being accessed by illegally.

Security and privacy are important issues when medical images and their patient information are stored and transmitted across public networks. The well-known standard for storage and exchange of medical images is DICOM (Digital Imaging and Communication in Medicine). A historical background and a brief overview of DICOM are reported in [1,2].

In steganography, the secret information (e.g. text, image, video, audio) one wishes to send is called the message. This message is embedded in a cover, which is typically an image, a video or an audio file. After the message is embedded, the cover becomes a stego. Here, we are focused on the scenario whereby both the message and the cover are 8-bit grey scale images. The reason for choosing images as a cover is images usually have a high degree of redundancy, which makes them suitable to embed information without degrading their visual quality. Moreover, images are widely exchanged over the internet than other digital media and they attract little suspicion.

General factors that need to be addressed by embedding techniques are: the quality of the stego, message detectability, payload capacity, and the robustness of the stego against distortion attacks. Least significant bit (LSB) replacement [3,4] and its variants are common embedding techniques in the spatial domain. Embedding in LSB of a pixel (i.e. only the LSB plane of the cover image is overwritten with the secret bit stream), makes it difficult to notice a change in the value of these pixel by naked eye. A review of few embedding techniques will be presented in Section 2.

In recent years, different Fibonacci systems of representing integers have been used as an alternative to the binary representation to improve quality and capacity by embedding in more than one bit-plane. In 8-bit grey scale images, a pixel value in the range 0-255 can be represented by 12 bits using Fibonacci representation [5]. However, unlike the binary based embedding methods, capacity of Fibonacci based embedding depends on the selected cover image -- not every pixel of the cover is a "good candidate" for the embedding. The non-uniqueness of Fibonacci representation of integers means that the embedding procedure cannot comply the Zeckendorf's theorem, i.e. the embedding procedure should not create a two consecutive 1s. To avoid this situation, Fibonacci based schemes may have to skip the current pixel and consider the next candidate pixel for selection [5]. In our paper, the redundancy problem of Fibonacci representation will be avoided the restrictions of Zeckendorf's theorem, and thereby resulting that each pixel of the cover can be used for embedding.

Whilst steganographers aim to design techniques to meet the goals of steganograpy, steganalysers attempt to defeat the goal of steganography by detecting the presence of a hidden message. There are a number of statistical and structural attacks to determine the presence/absence of a hidden message and estimate the size of the embedded secret message. Two well-known steganalysis techniques used to detect and estimate the secret message are Regular Singular groups (RS) [6], and Weighted Stego (WS) [7].

Some steganographic schemes attempt to be robust against steganalysis while others attempt to have better stego quality, but such schemes usually have capacity limitations [8]. Decreasing the capacity increases the un-detectability [9]. There is a trade-off between hiding capacity and quality of stego image. It is a challenge for a steganographer to achieve a good balance among the different steganography requirements. It is worth noting that is no agreed benchmarking standard to evaluate steganography algorithms [10]. An international standard for evaluating the security, performance, and effective-ness of steganography algorithms would be of great benefit to security researchers, and to producers and consumers of steganography products.

This paper is primarily concerned with improving quality and robustness aspects of steganography techniques whilst maintaining a full payload capacity. The contributions of the paper are: 1) we present an algorithm, SISR, to reduce the secret image size as a pre-processing step prior to secret image embedding; and 2) we present a bit-plane mapping technique for Fibonacci based message embedding. As a result of these two proposals, we are able to improve stego image quality and un-detectability without affecting the payload capacity. Although our focus for SISR algorithm is to reduce embedding impact on the stego image, it can also be applied on the well-known DICOM standard images for storage and exchange of medical images. The proposed SISR algorithm is useful for both secure and privacy-preserving archiving/storing and transmission of medical images. Furthermore, the output bits from SISR algorithm that represent the DICOM image can be embedded with less impact on the stego quality.

The rest of the paper is organized as follows: a review of literature on steganography is presented in Section 2. Section 3 presents the proposed SISR algorithm to pre-process secret images before embedding and Section 4 presents the proposed embedding technique. The experimental results are shown in Section 5. Finally, our conclusions and direction future work are given in Section 6.

## 2    Literature Review

The LSB [3,4], LSB Matching (LSBM) [11], and LSBM Revisited (LSBMR) [12] are basic embedding techniques. Other embedding techniques based on region selection such as randomly embedding across the cover [13], and embedding in edge areas [14,15,16] have been proposed. The sequential LSB replacement increases even pixel values either by one or they are left unchanged, while odd pixel values are decreased by one or are left unchanged. This could create an imbalance in the embedding distortion, referred to in the literature as the *asymmetry* problem, in the stego image that can be exploited to detect the presence of a hidden message. More seriously, the secret message embedded by the sequential LSB replacement is easily detected. This problem can be partially mitigated by the use of pseudorandom number generator (PRNG) to randomly distribute the message across the cover instead of embedding the message in sequential order [13]. This is called LSB randomly embedding. However, the LSB randomly still suffers from the asymmetry problem.

The LSBM scheme is designed to solve the asymmetry problem by checking if the next secret bit does or does not match the LSB of the corresponding cover pixel. If not, then LSBM randomly increases or decrease the pixel value by 1 [11]. LSBM does not suffer from the asymmetry problem, and has the advantage of high payload capacity, as well as good visual imperceptibility property [17]. In these LSB methods, the probability of pixel value change is 0.5 (i.e. they have almost the same stego quality). Mielikainen [12], used a function to develop the LSBM technique called LSB Matching Revisited (LSBMR) to improve the imperceptibility of the stego whereby two secret bits are embedded in a pair of pixels of the cover. Incorporating this function reduces the probability of changing pixel values from 0.5 to 0.375. However, these improvements come at the expense of limited payload capacity because LSBMR algorithm cannot be performed on saturated pixels.

Another approach to secret hiding is to embed a message in selective regions of an image using some criteria. Embedding the secret message in edge area of the cover helps improve imperceptibility and robustness against detection [14]. Sobel, Prewitt, and Canny [15] are the most popular edge detection techniques that have been used in steganography. However, edge based embedding have problems with determining the same edge area by the receiver because the act of embedding in an edge area could change the original edge pixels into non-edge pixels. Therefore, some parts of a secret message may be lost, i.e. the survivability of secret message cannot be guaranteed at the receiving end. Hempstalk, [14], suggested to overcome this problem by first replacing all LSBs of the cover with 0. Then the edge detection technique is applied on the modified cover to select the edge pixels for embedding. The receiver applies the same pre-processing then use the edge detection technique on the modified stego to select edge pixels. In [16], Weiqi et al. attempted to tackle the same problem by limiting the edge pixels involved in embedding using a threshold that depends on size of the secret message, and proposed an embedding algorithm based on edge areas and LSBMR embedding method. Their algorithm is, therefore adaptive, and only sharper edge pixels are used for embedding. All selective regions (variants of edge-based) algorithms have the advantage of improved quality of stego and detectability, but at the expense of limiting payload capacity because the secret bits can be embedded in only edge areas.

Fibonacci bit-plane representation of the cover pixel has been investigated for steganography due to their advantages over binary representation when embedding in higher bit-planes than the LSB. This alternative representation of integers provides more bit-planes than the binary [18]. When embedding in higher bit-planes, the effect on the quality of the stego is less with the Fibonacci than with the binary [5].

Diego et al. [5], produced a Fibonacci embedding technique by strict observation of the Zeckendorf theorem. In their scheme, they first select a pixel then decompose the pixel value into Fibonacci and also select the plane that use for embedding. Then they check if the selected pixel is a good candidate or not, and if it is not, skip it and select the next candidate pixel. If it is a good candidate, then the secret bit is replaced with the agreed bit-plane. They claimed that the same embedding scheme can be also applied to different planes resulting in more robust data hiding and possibly higher visual distortion. As they mentioned, the main aim of their scheme is to investigate the possibility of inserting a secret bit without altering the perceptual quality of stego. Also they claimed that if the secret bits are embedded in the LSB using binary or Fibonacci based embedding, the PSNR of Fibonacci becomes similar or higher comparing to the binary embedding based. The limitation of their algorithm is, not every pixel of the cover is going to be used for embedding.

A generalized Fibonacci decomposition is proposed in [18] as an improvement of [5]. The most common generalization of Fibonacci is the p-number Fibonacci sequence. In order to provide more places for embedding, the authors of [18] investigat-ed the p-number Fibonacci bit-planes to determine which planes are suitable for embedding. In their scheme, they first decompose the selected pixel into bit-planes using p-number Fibonacci. Then the selected plane is chosen for embedding and also the Zeckendorf theorem is applied on the selected pixel. Finally, they did a comparison between the proposed scheme and binary embedding in term of quality and capacity. As a result, even they claimed that the proposed scheme is better than binary in term of capacity and quality, but still their limitation is capacity because some pixel values are not candidates for embedding.

The use of more advanced versions of Fibonacci, and other representations have been investigated [19,20,21] to increase security. Fibonacci-like steganography by bit-plane(s) mapping instead of bit-plane(s) replacement has been proposed to increase embedding capacity by embedding two bits of the message in three bits of a cover pixel. However, the increase of security is achieved at the expense of a marginal loss in stego quality. In this paper, we extend the work in [21] to improve cover quality by embedding one secret bit in a cover pixel whilst maintaining the same robustness against statistical steganalysers. In this paper, the purpose of using Fibonacci representation instead of the binary representation is to gain higher stego quality, i.e. by embedding the obtained bits from SISR algorithm into Fibonacci representation of a cover pixel provides higher PSNR than embedding into binary representation (see Tables 9,11, and 12). This happens because of two reasons, first the obtained stream of bits from the proposed SISR algorithm that represent the secret image contains more 0s than 1s (see Tables 7 and 8) for natural images, the reason is discussed in Section 5.3. The second reason is in the Fibonacci representation complying with Zeckendorf theorem, more cover pixels' LSB have a value of 0 than 1, while in binary representation the number of cover pixels' LSB that have 0 value are almost equal to those have 1 (see Table 10). These two reasons increase the probabilities of similarity/matching between the secret bit and the cover pixel's LSB of the Fibonacci representation. Moreover, unlike the exiting Fibonacci-based bit-plane replacement techniques, the proposed mapping algorithm can be applied on all pixel intensities to embed secret bits.

## 3   Secret Image Size Reduction (SISR) Algorithm

Here we propose the SISR algorithm to reduce the size of secret images prior to embedding it in a cover image in order to reduce the impact of embedding on the stego quality (i.e. PSNR).

### 3.1 SISR Algorithm: Encoding

The SISR encoding algorithm works as follows:

- Split the secret image into non-overlapping blocks of size 4x4.
- Let $B_{ij}$ be a block of 16 pixels, and i, j ∈ {1,.., 4}. For each such block do the following steps:

1. Let $m = \min_{ij}(B_{ij})$, be the minimum pixel value, and let i*, j* be the indices of the element in $B_{ij}$ achieving $m$ with smallest j, then smallest i.
2. Let $D_{ij} = B_{ij} - m$, be the difference between each pixel and $m$.
3. Let $D_{max} = \max_{ij}(D_{ij})$, be the maximum difference value.
4. Let T be a set of possible thresholds to determine the length of bits that represent $D_{ij}$.

$$T = \{2^n - 1 \mid 0 \leq n \leq 7\}$$

5. Let $T^* = \min(z)$, where $z \in T$ and $z \geq D_{max}$, in other words, $T^*$ is the closest value in T to $D_{max}$ that is greater than or equal to $D_{max}$.

When $T^* = 0$, it means that the maximum value of differences $D_{ij}$ between each pixel and the minimum pixel ($m$) is 0, i.e., all pixels in that block have the same value. When $T^* = 1$, it means that the maximum value of differences $D_{ij}$ between each pixel and the minimum pixel ($m$) is 1, and so on for other values of $T^*$.

6. If $T^* = 255$, record 1 and followed by the original pixels' value in 8 bits, otherwise record 0. If $T^* = 0$, the recorded 0 is followed by the minimum pixel value in 8 bits, then the value of $T^*$ in 3 bits (111), and no information is needed to represent the position of the minimum pixel value because all of the pixels have the same value, therefore in this case only 12 bits are needed to represent the block. Otherwise, when $T^* = 1, 3, 7, 15, 31, 63$ or $127$, the recorded 0 is followed by:

   — 8 bits: indicates the minimum pixel value.
   — 3 bits: indicates the value of $T^*$, should be 110, 101, 011, 100, 010, 001, or 000 representing 1, 3, 7, 15, 31, 63, or 127 respectively.
   — 4 bits: indicates as the position of the minimum pixel value (in each block there are 16 pixels) therefore the positions are from 0 to 15.
   — From $T^* = 2^L - 1$, L is the length of bits that are needed for representing $D_{ij}$, where $1 \leq L \leq 7$

Note that the 8 bits that indicate the minimum pixel value and the 3 bits that indicate the value of $T^*$ are needed when the image pixel value is between 0 and 255. It is possible to extend this reduction algorithm to be applicable on different pixel value ranges, for example if the image pixel value is between 0 to 511, we can extend the algorithm to be applicable by representing the minimum pixel value in 9 bits instead of 8 bits and the value of $T^*$ in 4 bits instead 3 bits.

### 3.2 SISR Algorithm: Recovery

The following steps describe the recovery of the original images from the obtained stream of bits $b_i$ described in Section 3.1:

1. Take the first bit; if it is 1 take each of the following 8 bits and convert them into decimal (repeat it 16 times). If this is the case, terminate here, otherwise follow the next steps.

2. Next 8 bits indicate *m*.
3. Next 3 bits indicate T*.
4. Next 4 bits indicate index of *m*.
5. From value of T*, the length of the bits are decided so as to represent $D_{ij}$. For example if T* = 15, 4 bits are needed to represent $D_{ij}$.
6. $B_{ij} = D_{ij} + m$.

### 3.3 Example Application of SIRS Algorithm

The example 4x4 block of pixel intensities in Table 1 will be used to illustrate the SISR encoding and recovery steps.

**Table 1**. Block of 16 pixels

| 30 | 25 | 26 | 35 |
|----|----|----|----|
| 35 | 22 | 29 | 28 |
| 31 | 24 | 22 | 29 |
| 30 | 34 | 32 | 30 |

**SISR encoding steps**

1. The minimum pixel value is 22 (00010110 in 8 bits), and its index is 5 (0101 in 4 bits).
2. The differences between pixels and the minimum pixel are presented in Table 2.
3. The maximum value of the subtraction, see 3$^{rd}$ and 7$^{th}$ column of the Table 2, is 13.
4. The nearest value in set T that should be equal or greater than the maximum value of the subtraction, which is 13, is 15 (i.e. T* = 15).
5. Now this stream of bits represents the block of Table 1:

   0: indicates that the algorithm has been done on the block, i.e. T* is not equal to 255.
   00010110: represents the minimum pixel value.
   100: represents the value of T*.
   0101: represents the index of the minimum pixel value.
   1000, 0011, 0100, 1101, 1101, 0111, 0110, 1001, 0010, 0000, 0111, 1000, 1100, 1010, 1000: representing values of difference (see 3$^{rd}$ and 7$^{th}$ column of Table 2) 8, 3, 4, 13, 13, 7, 6, 9, 2, 0, 7, 8, 12, 10, 8 respectively.
   Finally, this stream (0 00010110 100 0101 1000 0011 0100 1101 1101 0111 0110 1001 0010 0000 0111 1000 1100 1010 1000) represents the 4x4 block (see Table 1), i.e. 76 bits are representing original 128 bits.

Table 2. Differences between pixels and minimum pixel value

| $B_{ij}$ | m | $D_{ij}$ | b | $B_{ij}$ | m | $D_{ij}$ | b |
|---|---|---|---|---|---|---|---|
| 30 | 22 | 8 | 1000 | 24 | 22 | 2 | 0010 |
| 25 | 22 | 3 | 0011 | 22 | 22 | 0 | 0000 |
| 26 | 22 | 4 | 0100 | 29 | 22 | 7 | 0111 |
| 35 | 22 | 13 | 1101 | 30 | 22 | 8 | 1000 |
| 35 | 22 | 13 | 1101 | 34 | 22 | 12 | 1100 |
| 29 | 22 | 7 | 0111 | 32 | 22 | 10 | 1010 |
| 28 | 22 | 6 | 0110 | 30 | 22 | 8 | 1000 |
| 31 | 22 | 9 | 1001 | | | | |

In Table 2, the 1$^{st}$ and 5$^{th}$ column, $B_{ij}$, is the values of the pixels in the Table 1 (excluding first minimum pixel value 22). The 2$^{nd}$ and 6$^{th}$ column is the minimum pixel value $m$, and the 3$^{rd}$ and 7$^{th}$ column, $D_{ij}$, is the subtraction of the minimum pixel value from each pixel value. The 4$^{th}$ and 8$^{th}$ column, b, represents the result of subtraction in binary form.

**SISR recovery steps**

From the obtained stream above, the original 4x4 block of pixels can be recovered as follows:

1. Take the first bit; which is 0, then go to the next step.
2. Convert the next 8 bits into decimal, which is 22, that represents the minimum pixel value $m$.
3. The next 3 bits, 100, represent the value of $T^*$, i.e. $T^*=15$.
4. The next 4 bits, 0101, represent the index of $m$.
5. As $T^* = 15$, take each next 4 bits 15 times and convert them into decimal to represent $D_{ij}$ as illustrated in Table 3.
6. Add $m$ to $D_{ij}$, then the original pixels $B_{ij}$ are obtained (see 4$^{th}$ and 8$^{th}$ column of Table 3).
7. Sequentially insert each value in the 4$^{th}$ and 8$^{th}$ column of Table 3 to its position in the block; the block is recovered exactly as it is (see Table 1).

Table 3. Producing original pixels from the recovered $D_{ij}$

| b | $D_{ij}$ | m | $B_{ij}$ | b | $D_{ij}$ | m | $B_{ij}$ |
|---|---|---|---|---|---|---|---|
| 1000 | 8 | 22 | 30 | 0010 | 2 | 22 | 24 |
| 0011 | 3 | 22 | 25 | 0000 | 0 | 22 | 22 |
| 0100 | 4 | 22 | 26 | 0111 | 7 | 22 | 29 |
| 1101 | 13 | 22 | 35 | 1000 | 8 | 22 | 30 |
| 1101 | 13 | 22 | 35 | 1100 | 12 | 22 | 34 |
| 0111 | 7 | 22 | 29 | 1010 | 10 | 22 | 32 |
| 0110 | 6 | 22 | 28 | 1000 | 8 | 22 | 30 |
| 1001 | 9 | 22 | 31 | | | | |

**Table 4.** Number of obtained bits from proposed SISR algorithm for block size 4x4

| T* | Number of bits | | |
|---|---|---|---|
| | Obtained | Original | Reduction |
| 0 | 12 | 128 | 116 |
| 1 | 31 | 128 | 97 |
| 3 | 46 | 128 | 82 |
| 7 | 61 | 128 | 67 |
| 15 | 76 | 128 | 52 |
| 31 | 91 | 128 | 37 |
| 63 | 106 | 128 | 22 |
| 127 | 121 | 128 | 7 |
| 255 | 129 | 128 | -1 |

Table 4 illustrates the number of obtained bits and the number of reduced bits depending on the value of T* for the 4x4 block. From this table, one can notice that only in the case T* = 255, the algorithm increases the number of bits by 1; otherwise, the algorithm reduces the number of bits to represent the block of 16 pixels.

Although the main focus in this paper of proposing SISR algorithm is to reduce the secret image size prior to embedding, it can also be used for the image storage and transmission purposes.

## 4 Proposed Embedding Algorithm

A PRNG is used to randomly select a cover pixel. Converting the cover pixel value into Fibonacci results in more than one representation for each pixel value. Complying with Zeckendorf theorem means that the first three LSBs of a cover pixel in Fibonacci representation belong to the set {000, 001, 010, 100, 101} as illustrated in Table 5. Depending on the set, our mapping embeds a secret bit into a cover pixel by mapping the secret bit onto the first 3 LSB bits according to Table 5.

**Table 5.** Mapping Algorithm

| 3 Cover bits | Secret bits | |
|---|---|---|
| | 0 | 1 |
| 000 | 000 | 001 |
| 001 | 000 | 001 |
| 010 | 010 | 001 |
| 100 | 100 | 101 |
| 101 | 100 | 101 |

From the structure of Table 5, the secret bits can be recovered by extracting the 1$^{st}$ LSB of the selected stego pixel using the same PRNG used at the embedding stage.

## 5     Experimental Results

Two sets of experiments were performed to evaluate the proposed algorithms: one to evaluate the proposed SISR algorithm and one to evaluate the performance of the proposed mapping based embedding technique.

### 5.1     Experimental setup

The following databases were used in our experiments:
1. The Miscellaneous volume of Signal and Image Processing Institute (SIPI) database of University of Southern California [22]. This database consists of 44 images of which 16 are color and 28 are monochrome images. We resized these 44 images to 512 x 512 and convert them into gray level images with 8 bits per pixel. These images will be considered as cover images. Also, the original 44 images were resized to 256 x 128 and converted to gray scale with 8 bits per pixel to be used as secret images. The reason for resizing these images to 256 x 128 is to make the number of bits that represent a secret image is equal to the number of cover pixels (262144 pixels).
2. Two sample databases of DICOM images: BU001015-V01 database contains 193 gray level images of size 256 x 256 with 16 bits per pixel, and MR database contains 140 gray level images of sizes 256 x 256 or 512 x 512 with 16 bits per pixel [23]. These two databases are used to test the proposed SISR algorithm.

### 5.2     SISR Algorithm Evaluation

The proposed SISR algorithm was applied on SIPI database images of size 256 x 128, BU001015 database images of size 256 x 256, and MR database images of sizes 256 x 256 and 512 x 512. The original number of bits in each tested image is 128 *256 *8 = 262144 bits, 256 x 256 x 16 = 1048576 bits, and 1048576 bits or 512 x 512 x 16 = 4194304 bits for SIPI, BU001015, and MR database respectively. We used the reduction ratio factor RR in Eq. 1 to evaluate the reduction efficiency of our approach.

$$RR = \frac{Total\ size\ in\ bits\ of\ the\ obtained\ bitstream}{Total\ size\ in\ bits\ of\ the\ input\ image} \qquad (1)$$

Table 6 shows RR after applying the SISR algorithm on the three databases for 3 different block sizes (i.e., 4x4, 8x8, and 16x16). From Table 6, we can see that the best RR is achieved with 4x4 blocks. The RR decreases as the block size increases. Furthermore, Table 7 and 8 demonstrates that the SISR algorithm produces a higher ratio of 0s bits than 1s.

**Table 6.** Average RRs for SISR algorithm for different block sizes

| Databases | Block Size | | | | | |
|---|---|---|---|---|---|---|
| | 4x4 | | 8x8 | | 16x16 | |
| | Average | Std. | Average | Std. | Average | Std. |
| SIPI | 0.697 | 0.141 | 0.754 | 0.138 | 0.841 | 0.111 |
| DICOM( BU001015) | 0.442 | 0.017 | 0.446 | 0.020 | 0.469 | 0.020 |
| DICOM(MR) | 0.426 | 0.031 | 0.429 | 0.034 | 0.455 | 0.035 |

**Table 7.** Ratio of Zero and One value bits obtained after applying the proposed SISR algorithm on 4x4block size

| Databases | Ratio of 0s | | Ratio of 1s | |
|---|---|---|---|---|
| | Average | Std. | Average | Std. |
| SIPI | 0.55 | 0.037 | 0.45 | 0.037 |
| DICOM( BU001015) | 0.58 | 0.004 | 0.42 | 0.004 |
| DICOM(MR) | 0.58 | 0.010 | 0.42 | 0.010 |

In order to embed DICOM images, we resized the first 10 images of databases BU001015 and MR to 128 x 128 to embed them in the first 10 cover images of SIPI database. The images were resized to 128 x 128 to make the number of bits that represent secret medical DICOM images equal to the number of cover pixels (262144 pixels). The result of the SISR algorithm is illustrated in Table 8.

**Table 8.** RRs and the ratio of Zero and One value bits Obtained after applying the proposed SISR algorithm on block size 4x4 for resized DICOM database images

| Databases | RR | | Ratio of 0s | | Ratio of 1s | |
|---|---|---|---|---|---|---|
| | Average | Std. | Average | Std. | Average | Std. |
| BU001015 | 0.403 | 0.038 | 0.60 | 0.005 | 0.40 | 0.005 |
| MR | 0.438 | 0.028 | 0.58 | 0.005 | 0.42 | 0.005 |

### 5.3 Embedding Algorithm Evaluation

Two experiments are conducted to evaluate the performance of the proposed embedding technique. The first is to test stego quality and the other is to measure the detectability of secrets. The results are compared with the binary based LSB randomly (LSB binary) and Fibonacci based LSB randomly (LSB Fibonacci) embedding technique. Images of SIPI database (44 images) of size 512 x 512 are used as cover images. For each cover image, 44 secret images of SIPI database of size 256 x 128 are embedded producing (44 x 44 = 1936) stego images for each tested embedding technique including the proposed embedding technique.

**Stego Quality.** Average PSNR values for the 1936 stego images are shown in Table 9 for LSB binary, LSB Fibonacci and the proposed embedding technique.

Table 9. Stego quality (PSNR) after embedding the original image (SIPI) vs. the reduced image

|  | Binary LSB | | Fibonacci LSB | | Fibonacci mapping | |
| --- | --- | --- | --- | --- | --- | --- |
|  | Original | Reduced | Original | Reduced | Original | Reduced |
| PSNR | 51.15 | 52.80 | 52.24 | 52.92 | 51.14 | 52.90 |

The second column of Table 9 presents the results when the original image bits are embedded in the cover using binary based LSB randomly embedding technique. The third column presents the results when reduced image bits are embedded using binary based LSB randomly embedding technique. The same for the fourth and fifth columns using Fibonacci based embedding technique. Finally, the sixth and seventh columns present the results when original image bits and reduced image bits are embedded using the proposed mapping based embedding technique. Results show that reducing the secret image using the proposed SISR algorithm leads to better stego quality compared to embedding the original image. The reason why the Fibonacci based LSB has higher PSNR value than others is because in the case of Fibonacci based LSB embedding technique, not every pixel is used for embedding, i.e. some pixels are excluded from embedding whereas in the binary based LSB and the proposed Fibonacci based mapping embedding technique, every pixel can be, and is, used for embedding. The average of maximum number of pixels, for the SIPI database, used for embedding for each tested technique is illustrated in figure 1. Moreover, the stego quality is better with the proposed mapping based embedding compared to binary based LSB. This is achieved because: 1) the proposed SISR algorithm always provide a bit stream that contains more 0s than 1s; and 2) in the Fibonacci representation compliant with the Zeckendorf theorem, the probability of pixels that have a value of 0 as the LSB is always greater than those that have 1. Whilst in the binary representation, the probability of pixels' LSB that have a value of 0 is equal to those that have 1. These two factors increase the probability of the secret bit and the LSB of the Fibonacci representation being the same. The reasons of why the proposed reduction algorithm provides a stream of bits that always contains more 0s than 1s are:

1. First recorded bit (see Section 3.1) is 1 when $T^*=255$ otherwise it is 0. As in most blocks of the secret image T* is not equal to 255, the stream has more number of 0s than 1s.
2. There is one possible 3-bit string (i.e., 000, 001, 010, 100, 101, 011, 110) to represent each T* value. The four most frequently occurring T* values are represented by the 3 bits with two or more 0s. Then the three remaining 3-bit strings represents the three least frequent T* value.

**Table 10.** Ratio of Zeros and Ones of cover pixels' LSB value for SIPI database

|  | LSB of Binary | | LSB of Fibonacci | |
|---|---|---|---|---|
|  | Number of 0s | Number of 1s | Number of 0s | Number of 1s |
| Average | 0.47 | 0.53 | 0.60 | 0.40 |
| Std. | 0.095 | 0.095 | 0.126 | 0.126 |

**Table 11.** Stego quality (PSNR) after embedding the original DICOM (BU00105) image vs. the reduced image

|  | Binary LSB | | Fibonacci LSB | | Proposed | |
|---|---|---|---|---|---|---|
|  | Original | Reduced | Original | Reduced | Original | Reduced |
| PSNR | 51.14 | 55.10 | 52.30 | 55.09 | 51.81 | 55.30 |

**Table 12.** Stego quality (PSNR) after embedding the original DICOM (MR) image vs. the reduced image

|  | Binary LSB | | Fibonacci LSB | | Proposed | |
|---|---|---|---|---|---|---|
|  | Original | Reduced | Original | Reduced | Original | Reduced |
| PSNR | 51.14 | 54.73 | 52.30 | 54.72 | 51.70 | 54.90 |

In Table 11 and 12, the first 10 images of databases BU001015 and MR are resized to 128 x 128 and embedded in each of the first 10 cover images of size 512 x 512 of SIPI database. Results in each cell are the average of 100 stego images. For the same reasons that we discussed before, we can see that the proposed embedding technique has higher quality than binary based LSB randomly embedding and Fibonacci LSB randomly embedding, and for the same reasons discussed before, the Fibonacci LSB has higher quality than binary based LSB randomly embedding and proposed embedding technique when the original secret image is embedded.

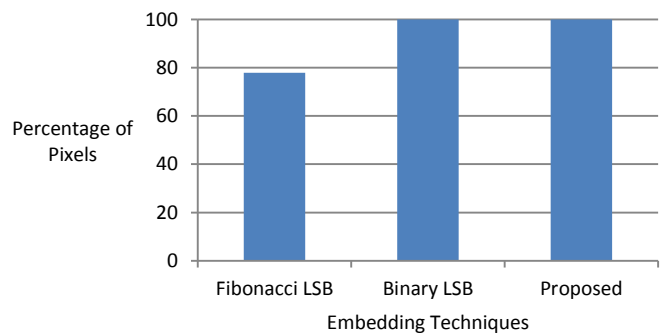

**Fig. 1.** Embedding Capacity

From figure 1, we can see that the Binary LSB and proposed embedding techniques have 100% of embedding capacity, i.e. every cover pixel can be used to embed the secret; while in Fibonacci LSB, only 78% of the cover pixels on average can be used for embedding the secret.

**Detectability.** Two steganalysers have been used to evaluate the detectability of the proposed embedding technique. Results are compared with binary based LSB randomly embedding technique.

In Table 13 and 14, binary LSB (Original) means the secret image is embedded using binary based LSB randomly; Binary LSB (Reduced) means the obtained bits from Table 6 ($2^{nd}$ column) is embedded using binary based LSB randomly; and proposed technique means the obtained bits from Table 6 ($2^{nd}$ column) is embedded using the proposed embedding technique.

*Regular and Singular (RS) steganalyser [6].* Fridrich et al. found that the RS ratio of a typical image should satisfy the rule: (RM $\cong$ RM- and SM $\cong$ SM-) through large number of experiments. When LSB of the cover is changed, the difference between RM and RM- and the difference between SM and SM- increases. Then, the rule is violated; therefore, one could conclude that the tested image carries a secret message. Note that when payload capacity p = 0 %, i.e. cover without an embedded message, the value of RM is close to RM-, and the value of SM is close to SM-. Depending on the description above, we can discuss the results in Table 13, which presents the average values of RM, RM-, SM, and SM- based on embedding 44 secret images of the SIPI database in each of 44 different covers of SIPI database resulting 1936 stego images. In the case of using binary LSB, embedding higher rate of secret bits lead to an increase in the differences between RM and RM-, SM and SM-, indicating the presence of a secret message. Whereas in the case of the proposed embedding technique, there are no such differences. This indicates that these images are non-stegos. Therefore, the proposed technique is robust against the RS steganalyser. When reduced-size messages are embedded using binary LSB, the RS is still able to detect the secret bits but with lower message size estimation. When embedding the reduced bits using the proposed mapping scheme, the RS is unable to detect the secret bits. This demonstrates that proposed embedding technique is robust against RS.

**Table 13.** Message detection based on RS steganalyser

|  | RM | SM | RM- | SM- |
|---|---|---|---|---|
| Binary LSB (Original) | 25.69 | 25.68 | 55.42 | 14.21 |
| Binary LSB (Reduced) | 28.63 | 24.06 | 53.40 | 15.56 |
| Proposed technique | 39.02 | 19.82 | 43.42 | 19.48 |

*Weighted Stego (WS) steganalyser [7].* When an image is submitted to the WS, its output indicates the probability of having a hidden message. A negative value is treated as 0 and any number >1 is an indication of full 100% secret load. The average result of 1936 stegos is displayed for each embedding technique. Table 14 shows the estimation results of the secret message length of two steganographic techniques, including our proposed by this steganalyser. From Table 14, it is noticeable that the proposed scheme mostly scores a value close to zero indicating the absence of an embedded secret. On the other hand, when embedded using the binary LSB, the steganalyser detects the secret message with high probability. This demonstrates that proposed embedding technique is robust against WS.

**Table 14.** WS Steganalyser

|  | Message length |
|---|---|
| LSB random (original) | 0.970 |
| LSB random (Reduced) | 0.734 |
| Proposed technique | 0.151 |

## 6  Conclusion

A two-step efficient steganography scheme has been proposed to enhance stego quality and secret un-detectability. The first step of reducing the number bits required to represent a secret image (i.e., SISR) prior to embedding led to improved stego quality. The second step is the Fibonacci-based mapping technique to embed the message bits. This technique further improved the stego quality by embedding the obtained bits from the SISR algorithm in the cover by bit-plane(s) mapping instead of bit-plane(s) replacement. The proposed embedding mechanism outperforms the binary based LSB randomly embedding in two aspects: stego quality and robustness against steganalysers. Furthermore, this scheme overcomes the limitation imposed by the Zeckendorf theorem which improves capacity as well as the stego quality. The improvement in stego quality is the result of the combined effect of the SISR which results in secret message bit stream that has more 0s than 1s and the nature of the Fibonacci representation of cover pixel values which results in more 0s at the LSB than 1s. The experimental results demonstrated that the proposed embedding scheme is also secure against RS, and WS steganalyser attacks.

Our future work directions are: 1) investigate all existing pixel value decomposition techniques such as Prime, Natural, Lucas, Catalan in terms of providing higher ratio of cover pixels' LSB having the zero value and also propose a new pixel values decomposition technique that provides higher ratio of cover pixels' LSB having zero value; and 2) find a mechanism to improve the SISR algorithm in order to provide higher ratio of obtained bits that their values are zeros than 1s. Thus, these two factors are increase the probability of similarity between the secret bits and cover pixels' LSB results in better stego quality and less message detectability.